# Early Detection of Diabetic Retinopathy and Severity Scale Measurement: A Progressive Review & Scopes

*Asma Khatun, Sk Golam Sarowar Hossain

Department of Computer Science & Engineering, Aliah University, Kolkata, India

**Abstract:**
Early detection of diabetic retinopathy prevents visual loss and blindness of a human eye. Based on the types of feature extraction method used, DR detection method can be broadly classified as Deep Convolutional Neural Network (CNN) based and traditional feature extraction (machine learning) based. This paper presents a comprehensive survey of existing feature extraction methods based on Deep CNN and conventional feature extraction for DR detection. In addition to that, this paper focuses on the severity scale measurement of the DR detection and to the best of our knowledge this is the first survey paper which covers severity grading scale. It is also necessary to mention that this is the first study which reviews the proposed Deep CNN based method in the state of the art for DR detection methods. This study discovers that recently proposed deep learning based DR detection methods provides higher accuracy than existing traditional feature extraction methods in the literature and also useful in large scale datasets. However, deep learning based methods require GPU implementation to get the desirable output. The one of the other major finding of this paper is that there are no obvious standard severity scale detection criteria to measure the grading. Some used binary class while many other used multi stage class.

*Keywords: Diabetic Retinopathy; Deep Learning; Feature Extraction; Severity grading; Literature Review; Scopes & limitations*

## 1. Introduction

Retinopathy is a disease or a condition that affects the retina of a human eye. Retinopathy is formed by various associations independently such as diabetes, hypertension, coronary heart disease, and chronic kidney disease [1]. A retinopathy which causes from diabetes mellitus (or DM) is called as diabetes retinopathy or shortly DR.
DR, a risky and progressive disease creates various impairment in the retina of the human eye. These impairments are called lesion. Retinopathy is also considered as one of the mortality disease in older person [1,2] and the risk increased more if the person has diabetes with the history of clinical stroke [1]. According to a global metaanalysis study report in 2010, 1 in 3 (34.6%) had any form of DR in the US, Australia, Europe and Asia [3]. DR can be broadly classified as non-proliferative DR (NPDR) and proliferative DR (PDR). NPDR and PDR are mainly differentiated by the blood vessels in the retina. In NPDR blood vessels damage and pass fluid into the retina [4]. Whereas in PDR, new abnormal blood vessels grow in the retina [5]. PDR is more advanced stages of DR. If the DR reaches into PDR stage, it may cause total blindness [5]. The interest of the automating system is to prevent and treat the retina from damage and loss and hence the literatures in this research are considered for the detection of lesions in NPDR stages. The major types of lesions exists in a NPDR images are Microaneurysms (MA), Hemorrhages (HEM), Hard exudates (HE) and Soft exudates or Cotton Wool Spots (CWS). Hemorrhages are appeared as dot and blot types. Wheareas MA and HEM are considered as red lesions or sometimes these are referred as dark lesions and HE and CWS are considered as bright lesions [6]. MA and HEM are both are red in color and they can be differentiated by the shape and size. Among all, MAs and dot HEM are the first visible sign of DR. MAs are look like circular spot with sharp margins in less than 125 μm in dimension. Normally HEM are larger than MA [7]. It is although difficult to distinguish MA and dot HEM visually but blot HEM are irregular shape margin and even or unevenly densed [5, 8]. HE appears as small white or yellowish-white deposits with sharp margins and located in outer layer of retina [8]. The other symptoms in DR shows as venous beading (VB), neovascularization (growing of abnormal vessels) and intraretinal microvascular abnormalities (IRMAs). These are types of abnormalities of the blood vessels that occurred in the retina of the eye [1, 9-10]. Apart from those, the most severe, vision-threatening diabetic retinopathy are neovascularization (PDR) and diabetic macular edema (ME) which is macular thickness (area within 1DD from the center of Macula)[8]. Interested readers are referred to the [11] for the information on the causes of the lesions occurred in the retina of a human eye. In general, the ophthalmologists examine and grade NPDR into three stages: i.e., mild, moderate and severe depending on the location and occurrence of the lesions [6] exist in a retina. Fig. 1 (a) shows a color fundus image normal of retina along with its main components. It also shows examples from different categories of NPDR, Fig 1(b, c & d). In the literature various ways of severity grading category exist. Following are the standard category for the severity measurement based on the lesions exists in a retina [12]. This grading criteria is also international classification of DR scale developed by International Council of Opthalmology [3]:

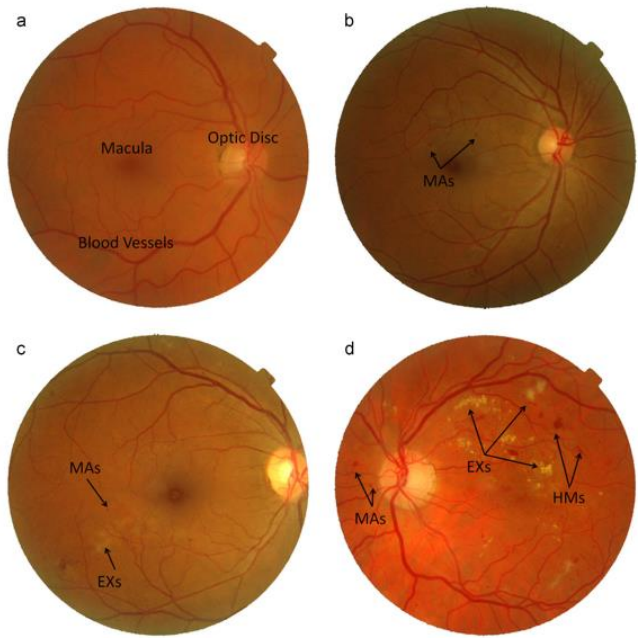

Fig. 1. Human retina and NPDR stages: (a) normal retina along with main components, (b) mild NPDR, (c) moderate NPDR, and (d) severe NPDR., images are adapted from Akram *et al.* [5]

**(i) Early/Mild NPDR** – At least one MA can be seen.
**(ii) Moderate NPDR** – There may be multiple MAs, dot-and-blot hemorrhages, venous beading, and/or cotton wool spots.
**(iii) Severe NPDR** – In the most severe stage of NPDR, CWS, venous beading, and severe intraretinal microvascular abnormalities (IRMA) can be found. It is diagnosed using the "4-2-1 rule." A diagnosis is made if the patient has any of the following: diffuse intraretinal hemorrhages and microaneurysms in 4 quadrants, venous beading in ≥2 quadrants, or IRMA in ≥1 quadrant. Within one year, 52-75% of patients falling into this category will progress to PDR [13].

Automatic grading based eye screening system is highly desirable replacing manual grading as manual grading system is time consuming and cost effective. This paper review several automatic DR detection techniques and explains current development on that. The main aim of this research is to review the current progress on automatic severity grading of the DR detection due to its necessity and efficiency in the clinical observation and treatment in timely manner. Google announces the three benefits of automatic grading as (i) increase efficiency & coverage of screening (ii) reduce accessing limits & (iii) improving patients' outcome by early detection [14]. Automatic grading based screening system is beneficial as the affected DR population is growing faster worldwide which will lead to difficulty for manual grader eye screening. Moreover, some researches show that the automatic grading system is cost-effective to the manual grading [15-21].

---

*Corresponding author (asma.sun03@gmail.com)

From the conducted study, it is found that very few research measure the severity to grade all the stages (multi-stage) using multiple lesions. But there are number of researches found to detect only the early stage (i.e., mild stage) by the presence of a single lesion such as MA or HE. Among all, the study which used deep CNN method are high discriminative in automated DR severity grading and also very useful in a large scale database. On the other hand, a noticeable research papers found to use the conventional feature vector ( machine learning classifier based) method for the automatic detection of DR and in recent, many of the severity grading researches are found to use feature vector based classification techniques. The advantage of the use of ML based classifier is that the automatic grading is possible easily by the assigned labeling of the extracted features for severity measurement. Furthermore, machine learning classifier become advance in respect to time complexity as well, a research with lower time complexity can be found in RoyChowdhuri *et al.* [22]. The variety of machine learning classifier such as Gaussian Mixture model (GMM), k-nearest neighbor (kNN) is used by the conventional approaches [5, 15, 22, 23]. However, these types of methods are not practical for high sensitivity (except the method proposed in [22] in the measure of sensitivity), high specificity and accuracy. Moreover, it is noteworthy to mention that the advancement of Graphics Processing Units (GPUs) has led to attract Deep CNN based feature extraction method for DR detection. Therefore several researchers implemented deep learning models for DR detection. Deep learning, an emerging powerful tool for DR detection, already exceeds high performance than the traditional feature extraction techniques. On that basis, the focus of this paper is on feature extraction methods based on deep CNN methods and traditional machine learning feature extraction based method. Besides those, this study looks for the researches of measuring performances by the use of cross validated training and testing datasets. Cross validation training and testing datasets has the potential to save time by not requiring the training samples [24]. A large number of significant research contributions on early detection of DR methods is available and there is very few review article were published such as [11, 25, 26 & 27]. The existing survey paper are categorized the method based on type of DR lesion detection like MA or exudates for example in [25-27]. However, there is none of these paper focuses on different grading for detection of DR. Compared with the existing literature on DR survey, the main contributions of this paper are as follows:

i. First and foremost, to the best of our knowledge, this is the first survey paper in the literature that covers the deep learning based feature extraction based method on DR detection.
ii. To the best of our knowledge, this is the first survey paper in the literature that focuses the survey papers in respect to the severity grading on different scaling of early detection of DR. Comparison is performed in regards to traditional feature extraction and deep CNN methods.
iii. As opposed to previous reviews, e.g., in [11] & [26], this article covers the most recent up to date literature of traditional feature extraction methods. A comparative summary in respect to the accuracy, severity grading stages performances of existing methods, application of database for severity scaled performance.
iv. This paper also provides insightful & critical analysis on the aspect of severity grading stage of the existing methods. It also finds future scope on deep CNN application. This paper covers the up-to-date review of research of the aforementioned application.
v. Apart from these, this study also gives attention to find the researches which measure their performances using cross validated training and testing datasets.
vi. Summarized performance of various methods is reported in tabular forms (e.g., Table 2- 4).

The rests of the paper is organized as follows. Section 2 describes the background concepts to the retinal imaging and acquisition, existing popular databases and their application to the severity grading. The section 2 also reported performance measurement tools applied in the literature. Comprehensive survey on the existing deep CNN, feature extraction (ML based) and the corresponding summary tables are provided in the Section 3. The severity scale measurement methods are discussed in the Section 4. Section 5 summarizes this study by discussing potential research scopes, and challenges and future works.

## 2. Background Concept

### 2.1 Data acquisition & camera concept:

Several methods of retinal image acquisition are available in the literature including slit-lamp biomicroscopy, nonmydriatic retinal photography, nonmydriatic retinal photography used with mydriasis, mydriatic retinal photography, fluorescein angiography and optical coherence tomograpgy (OCT). Among all, conventional fundus imaging is based on mostly mydriatic retinal photography examples include eyePACS [28], MESSIDOR [29], DRIVE [30]. Mydriatic and nonmydriatic both cameras have the advantage of variety of field view imaging capability. However, nonmydriatic cameras are less costly, produce clear and magnified image, digital recording, more comfort and safer compare to direct opthalmoscope [31]. On the other hand, fluorescein angiography or FA imaging are processed by intravenous injection of dye that increases the contrast of the blood vessels against the background [32]. This type of camera is not recommended for retinal screening due to the significant time required for the imaging. The high contrast structural OCT imaging has the advantage of quick and without dye needed technique compare to FA techniques. OCT imaging are shown to get promising and transformative technology but there still require further investigation and research to use widely [33].

### 2.2 Databases:

Many databases have been built to test various algorithms. The popular and publicly available databases are EYEPACS, MESSIDOR, DRIVE, STARE, and DIARETDB, KAGGLE. In recent study, some other database are used which are collected from Singapore National Diabetic Retinopathy Screening Program (SIDPR). Below Table I provided brief summary description with their major characteristics of the recent used databases for DR detection.

Table I: A comparative summary on the existing database

| Database | Information | Graded | Grading criteria |
|---|---|---|---|
| **Eyepacs [28]** | 3 million retinal images; Pupil dialation require only when pupil are small and not enough | Yes (DR & ME both) | 1: No DR, 2: MA only, 3: CWS, 4: HEM with or without MA, 5: VB, 6: IRMA, 7: Growing of new vessels |
| **Messidor 1 [29]** | 1200 retinal images; 66% Images captured by pupil dilation. | Yes (4 class) | 0: (Normal): (MA = 0) AND (HEM = 0)<br>1: (0 < MA <= 5) AND (HEM = 0)<br>2: ((5 < MA < 15) OR (0 < HEM < 5)) AND (NV = 0)<br>3: (MA >= 15) OR (HEM >=5) OR (NV = 1)<br>**Value are the no. of lesion presence** |
| **Messidor 2 [30]** | Total of 1748 image | Yes (2 class) | |
| **DRIVE [31]** | Total 40 color retinal images; | No | No |
| **STARE [35]** | 20 color retinal images | No | No |
| **ImageRet [36]** | DIARETDB0: 130 images<br>DIARETDB1: 89 images | No | No |
| **Kaggle [37]** | 35,000 datasets | Yes (5 class) | |
| **SIDPR [38]** | Total 494661 retinal images of approximate 170000 patients of variety camera setting; Multi ethnic data recorded; Telemedicine based screening program | Yes | By professional doctor |

### 2.3 Performance measurement:

The popular performance evaluation tools such as Sensitivity, Specificity, Accuracy, AUC are used for the DR detection severity grading. Those are briefly described below.

**Sensitivity (SN):** This metric is defined in terms of true positive and false negative. False negative is the case when the algorithm misclassified lesion as a non- lesion. True positive is the amount of the correct classification of the lesion detection.

$$SN = \frac{TP}{TP+FN} \tag{1}$$

**Specificity (SP):** This metric is defined in terms of true negative and false positive. False positive is the case which actually contains a lesion but algorithm unable to detect as a lesion. True negtive is the amount of correct classification of the non- lesion detection.

$$SP = \frac{TN}{TN+FP} \tag{2}$$

**Accuracy (Acc):** $Acc = \frac{TP+FN}{TP+TN+FP+FN}$ (3)

**The value of the area under ROC curve (AUC):** It is defined by how much system is sensitive to detect the desired output? The optimal measurement is 1.

## 3. Early Detection of DR: A Literature Survey:

In the literature, early detection method of DR can be broadly classified in two major classifications as: 1. Direct method (Deep CNN) based & 2. Traditional feature extraction based

(i) Direct Method (Implicit feature)
Direct methods are only seen to the use of deep convolution Neural Network architecture. In the literature ImageNet [39], AlexNet [40], GoogleNet [41] and Inception-V3 architecture [42] are seen to apply to training the DR images. These are direct method because they do not need to extract feature explicitly rather they implicitly learn the pattern of DR anomalies and provide the grading result according to the grading criteria. Moreover, in direct methods, there is no need to define feature vector. These types of methods are comparably newest research in the literature.

(ii) Explicit Feature extraction & Machine Learning
On the other hand a typical feature extraction based DR detection system consists of three main phases: preprocessing, feature extraction and classification. Preprocessing has several steps such as image segmentation, background pixel removal, image contrast enhancement etc. 'Optic disk (OD) and blood vessel' detections are the stages of image segmentation. OD detection is one of the major steps of image segmentation, otherwise system might detect OD as lesion. In the background pixel removal phase the surrounding black background pixel is removed from the retinal image to separate the retinal pixel for processing. Contrast enhancement techniques are used to enhance the image characteristics for clear lesion detection. During the feature extraction phase, feature descriptors are explicitly build up by the researchers. At the end machine learning classification approaches are used for DR grading. The both types of survey papers are described in the following sub sections.

### 3.1 Direct Methods

In the current development and recent history direct methods of using deep convolution Neural Network (CNN) or deep learning based method provides most highest accuracy of DR detection. In briefly, CNN is a type of multi-layer generative model that learns to extract meaningful features (high level feature) which resemble those found in the human visual cortex. In addition to that, a parallelization algorithm is also used to distribute the work among multiple machines connected on a network, and hence the training model can be done in reasonable time [43]. Deep learning based CNN architecture model such as ImageNet, GooGleNet, AlexNet are highly successful on image recognition task. Indeed, the authors of all those deep convolution NN architecture were build for the purpose of natural image recognition of both types of images which can be discriminated by human vision as well as those are not able to discriminate by human eye [44]. In the recent literature, these CNN based methods becomes very popular. Highly successful researches found in Lam *et al*. [45], Gulshan *et al*. 2016 [46] and Abramoff *et al*. 2016 [47] by the applying deep CNN such as GooGleNet, Inception, AlexNet architecture respectively. On the other hand, attempted works by the use of deep CNN architectures seen by Takahashi *et al*. 2017 [48], Pratt *et al*. 2016 [49] and Alban & Gilligan [42] did not show promising results. An example of a successful deep learning based DR detection method [50] and the architectural framework of corresponding convolution neural network (CNN) is presented in the Figure 2.

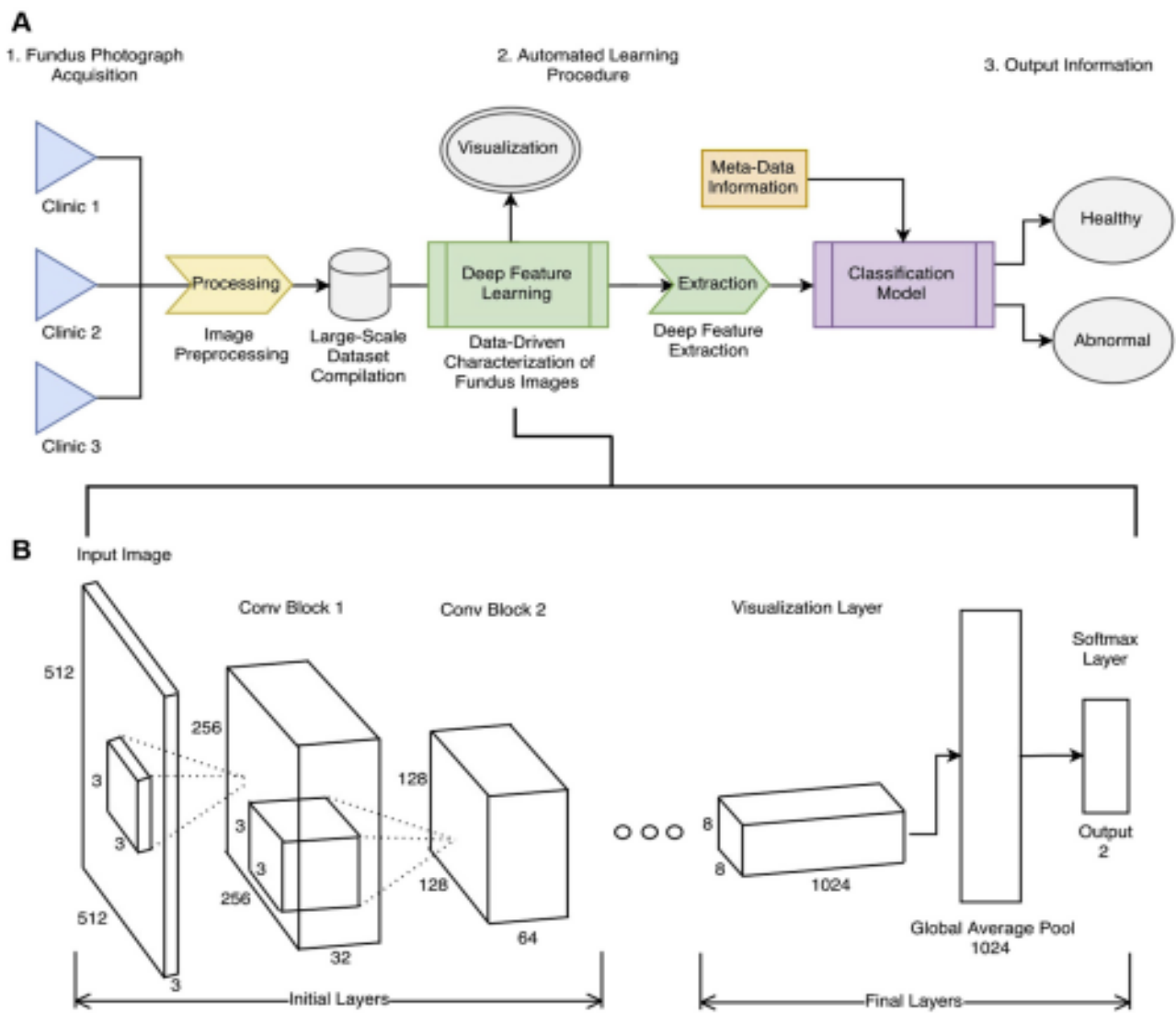

Fig. 2. Abstract view of the algorithmic pipeline proposed by Gargeya & Leng [50]. (**A**), Integration of the algorithm in a real diagnostic work flow. (**B**), Abstraction of the deep neural network. Features were extracted from the global average pool layer for a total of 1024 deep features. **Conv = convolutional**, Gargeya & Leng [50].

### 3.1.1. Deep learning based survey methods description and summary

In most recent studies, Lam *et al.* [45] explored his idea to use a deep CNN model of 22 layers called GoogLeNet. They mentioned in their report that more convolution layer allow the model to learn deeper features. This effective method has gained 95% SN, 96% SP in binary stage classification on Kaggle datasets. However, they discovered that poorer results occur in multistage classification with the same framework as the distinct features are undetected. After that they considered to changed the datasets from Kaggle to Messidor (for multistage) and noticed that noticeable results has increased which are shown in the Table V. Recently, in another proposed approach using enhanced deep network (AlexNet & oxford visual geometry group for training) by Abramoff *et al.* [47] reported higher sensitivity as 96.8%, 87% specificity & 0.98 of AUC by the use of MESSIDOR2 database. In another recent work, Gulshan *et al.* [43] presented a method of DR severity grading by the use of deep CNN, Inception-V3 architecture model which is an adapted model from Szegedy *et al.* [42]. The local features mainly neighborhood pixel and their aggregate is used as a pattern to learn the characteristics of various types of lesions such as MA or HEM during the training. They have highly successfully achieved 0.99 AUC for the testing of EyePACS-1 and MESSIDOR2 databases. The sensitivity of 90.3% and 87% were achieved in the EyePACS-1 and MESSIDOR2 databases respectively. 98.1% and 98.5% of specificity were achieved in the EyePACS-1 and MESSIDOR2 databases respectively. It is observed that many experimental works have implemented deep learning models for DR detection, which reaches similar performance or in some cases exceeding that of alternative techniques. Table II summarized the current development of the literature survey of DR detection based on Deep CNN approaches.

Table II: A short description on existing deep CNN based method

| Author | Results | Grading-stage | Methods Used | Preprocessing | Datasets | Year |
|---|---|---|---|---|---|---|
| Lam *et al.* [45] | 95% SN 96% SP | Binary class | Deep GoogLeNet | Histogram equalization | Kaggle & Messidor-1 | 2018 |
| Lam *et al.* [45] | SN 98%, 7% & 93% resp. | Multi class | Deep GoogLeNet | Histogram equalization | Kaggle | 2018 |
| Gargeya & Leng [50] | 0.97 AUC 94% SN 98% SP | Two class (DR & No DR) | Deep learning & added 3 metadata (1024+3) features with tree-based classifier for both training & validation | Scale, rotation & contrast invariant | EyePACS ( 5-fold cross validation) | 2017 |
|  | 0.94 AUC 93% SN 87% SP |  |  |  | Messidor- 2 |  |
|  | 0.95 AUC |  |  |  | E-Optha |  |
| Gulshan *et al.* [43] | 0.99 AUC 90.3% SN 98.1% SP | Five class | Deep learning; pixel neighborhood & aggregataion based predicted model | Scale normalization | EyePACS | 2016 |
|  | 0.99 AUC 87% SN 98.5% SP |  |  |  | Messidor-2 |  |
| Pratt *et al.* [49] | 75% Acc 95% SN 30% SP | Five class | Convolution NN using keras deep learning package | Color normalization | Kaggle | 2016 |
| Abramoff [47] | 0.98 AUC 98.6% SN 97% SP |  | Deep AlexNet | ---- | Messidor-2 | 2016 |
| Takahashi [48] | 81% Acc | Four | Deep GoogLeNet | ---- | Own-single field | 2017 |
| Alban & Gilligan [51] | 0.79 AUC 45% Acc | Five | GoogLeNet, AlexNet | Image denoising | EyePACS | 2016 |

### 3.2 Traditional Feature Extraction Methods Description and Summary

This section presents the existing literature survey based on traditional feature extraction method. There are three major steps in this type of feature extraction based methods and as follows: preprocessing, possible lesion detection (feature detection) and feature classification. The main methods in preprocessing are included as background detection, optic disk detection, contrast enhancement, blood vessel detection. This article is limited to the detail discussion of preprocessing methods. An example of traditional feature extraction method proposed by Akram *et al.* [52] is shown in the Figure 3 below.

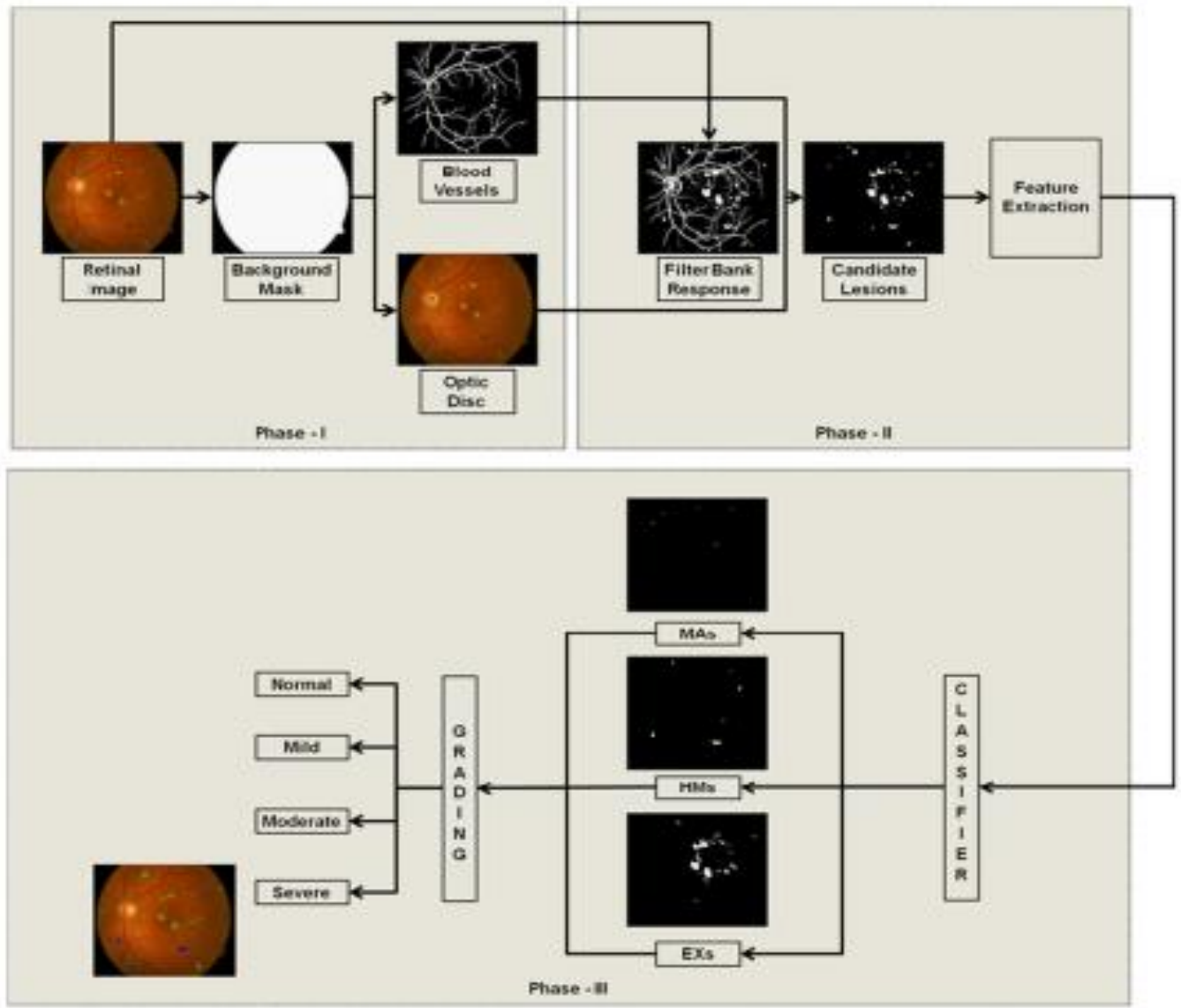

Fig. 3. A typical example of DR detection method proposed by Akram et al. [52]

By the use of 16-D feature vector and m-Mediods & Gaussian mixture model ensemble classification technique, Akram *et al.* [52] achieved 99.17%, 97.07% & 98.90% of image level SN, SP and Acc respectively for severity grading stages in MESSIDOR database. In the method of RoyChowdhuri *et al.* [53], 30 pixel based features are selected for classifying and grading the DR out of 70 features. The feature selection procedure was adapted by Cherkassky and Mullier [54] by the use of AdaBoost classifier. The range of features were considered from structural or regional pixels such as pixels area, convex area, orientation, standard deviances in object regions etc. GMM and kNN based classifier were used to classify the DR images and graded the severity based on the presence of MA and HA. They have succeed to achieve 100% sensitivity, 53.16% specificity, and 0.904 AUC by the use of MESSIDOR databases by the use of SVM & KNN classification method. Table III summarized the current development of the literature survey of DR detection based on traditional feature extraction approaches.

Table III: A short description on existing traditional feature extraction based method

| Author | Year | SN(%) | SP(%) | AUC | Acc | Classifier | Features | Database | Grading |
|---|---|---|---|---|---|---|---|---|---|
| Pires et al. [55] | 2015 | 100 | 88.9 | 0.97 | - | Bag of Visual Words (of Point of Interest) | Coding/pooling of visual words lesions attributes | Cross datasets training and testing (by their own build data) | Bright and dark multi-lesions |
| RoyChowdhuri et al. [53] | 2014 | 100 | 53 | 0.90 | | SVM+KNN | Pixel based 30 features selected using AdaBoost classifier, OD & blood vessel detection | MESSIDOR | Yes (0 vs 1, 2, 3)* |
| Akram et al. [5] | 2014 | 97 | 97 | 97 | | Hybrid classifier of GMM & SVM | Set of 16 features, background pixel removal, OD & blood vessel detection | MESSIDOR | Yes (lesion level) |
| Antal et al. [56] | 2012 | 96 | 51 | 0.87 | 0.90 | Ensembling of multiple classifier | gray level transformation, histogram equalization, MA enhancement etc. | MESSIDOR | DR/No DR |
| Fraz et al. [57] | 2012 | 75 | 97 | 0.97 | 95 | Ensemble classifier, decision tree based bagging and boosting | 9 features vector of Vessel line strength, Gabor filter and morphological transformation | STARE | bright and dark lesions |
| Fraz et al. [57] | 2012 | 74 | 98 | 0.97 | 94 | Ensemble classifier, decision tree based bagging and boosting | 9 features vector of Vessel line strength, Gabor filter and morphological transformation | DRIVE | bright and dark lesions |
| Fraz et al. [57] | 2012 | 72 | 97 | 0.96 | 94 | Same as above | Same as above | Cross datasets of trained with STARE and tested on DRIVE | bright and dark lesions |
| Esnaashari et al. [58] | 2011 | 95 | 89 | 91.86 | - | Neural network | Background removal contrast enhancement, gabor filter | MESSIDOR | DR at different stages & ME |
| Barriga et al. [59] | 2010 | 98 | 68 | 0.86 | | Partial least square, SVM | AM-FM features | MESSIDOR | DR/no DR |

## 4. Comparison Summary based on Severity Scale

Existing methods used mainly two measures as severity grading to analyze the statistical performance measurement. Some of the research categorized the severity grading only in two (binary 0 and 1) classes as 0 for no DR and 1 indicates DR exist in any severity (such as mild, moderate, severe and PDR). The above mentioned categorization seen in both CNN based and traditional feature extraction based method for measuring the performance (short name has been provided as TE for the sake of clarity when presenting in tabular form). In the Table IV, the comparison summary is presented based on severity grading of binary classification. Other measure is called multiclass grading scale such as mild, moderate and severe grading which are also taken into account by many researchers. Table V presents the comparison summary of the surveys considering multiclass severity scale.

Table IV: Performance on both CNN and TE methods on binary classes severity scale

| Method | Grading | Results | Database |
|---|---|---|---|
| Gargeya & Leng [50]$_{CNN}$ (2017) | 0 vs 1 | 94% SN, 98% SP, 0.97 AUC | EyePACS |
| | | 93% SN, 87% SP, 0.94 AUC | MESSIDOR2 |
| | | 90% SN, 94% SP, 0.95 AUC | E-Optha |
| Lam et al. [45]$_{CNN}$ (2018) | 0 vs 1 | 95% SN, 96% SP | Kaggle & Messidor-1 |
| Seoud et al. []$_{TE}$ (2016) | 0 vs 1 | 94% SN, 50% SP, 0.90 AUC | 6 publicly available databases |
| Roychowdhury et al. [53]$_{TE}$ (2014) | 0 vs 1 | 100% SN, 53% SP, 0.90 AUC | MESSIDOR |
| Antal et al. [56]$_{TE}$ (2014) | 0 vs 1 | 90% SN, 91% SP, 0.99 AUC | MESSIDOR |
| Barigga et al. [59]$_{TE}$ (2010) | Normal vs abnormal | 98% SN, 67% SP, 0.6 AUC | MESSIDOR |

Table V: Performance on both CNN and TE method on multiclass severity scale

| Method | Grading | Results | Database |
|---|---|---|---|
| D. SW Ting et al. [2017]$_{CNN}$ (2017) | moderate | 90.5% SN, 91.6% SP, 0.936 AUC | SIDPR |
| | vs severe | 100% SN, 91.1% SP, 0.958 AUC | |
| Lam et al. [45]$_{CNN}$ (2018) | 5 class, 3 arry | Acc of 74.5%, 68.8%, and 57.2% resp. | Kaggle |
| | No DR, mild vs severe DR | SN 98%, 7% & 93% resp. | Kaggle |
| | No DR, mild vs severe DR | SN 85%, 29% & 85% resp. | Messidor-1 |
| Gulsan et al. [43]$_{CNN}$ (2016) | moderate | 90.1% SN, 98.2% SP | EyePACS-1 |
| | severe | 84% SN, 98.8% SP | |
| | moderate | 86.6% SN, 98.4& SP | MESSIDOR-2 |
| | severe | 87.8% SN, 98.2% SP | |
| Abramoff et al. [47]$_{CNN}$ (2016) | multiclass | 96.8% SN, 87%.0 SP, 0.98 AUC | MESSIDOR-2 |
| Ensari | | | |
| Seoud et al. [60]$_{TE}$ (2016) | Mild vs severe | 96.2% SN, 50% SP, 0.916 AUC | 6 publicly available datasets |
| Barigga et al. [59]$_{TE}$ (2010) | Normal vs referral | 100% SN, 100% SP, 0.98 AUC | MESSIDOR |

From the above results it is observed that binary class severity grading based DR detection methods provides higher accuracies compared to multi- class detection. Moreover Lam et al. [45] reported that deep learning for binary classification in general has achieved high validation accuracies while multi-stage classification results are less impressive, particularly for early-stage disease. But in clinical practice, multi stage classification detection is more necessary. Hence, it requires more investigation on this paradigm.

## 5. Merits and Demits of the Algorithms

As discussed in the Section 3, it can be argue that the existing deep CNN methods gained (maximum of 98.6 % SN rate & 0.98 of AUC) most discriminating accuracy than the traditional feature extraction based methods. Below are discussed advantages and disadvantages of the existing methods in the literature.

- Traditional feature extraction based methods are defined to detect discriminating features such as different types of lesions (MA, HE, HEM, CWS) by explicitly (manually). Whereas in deep learning based methods avoid for defining such explicit feature extraction procedure, indeed these are involved to predict feature directly from the raw images by learning the pattern or level of the images from the datasets.

- However, traditional feature extraction methods are more suitable for small data scale. However, deep CNN perform better in the case of large scale datasets, hence computational cost increases. But, in clinical setting, data is often limited.

- On the other hand, deep learning based methods outperform conventional feature extraction approaches and it has rapid adaption to the DR detection due to several open source packages exists. Moreover, conventional methods already become old fashion. A number of experimental works have implemented deep learning models for DR detection, reaching similar performance or in many cases exceeding that of alternative techniques. However, none of the approaches mentioned the reason for choosing CNN model architecture, hence the question raised for datasets and software integration. Even most of the researchers can not provide the reason of good results or any modification requirement.

- In addition to that, to control the CNN architecture, many parameters such as size, number of filters are required, which is quite cumbersome technique. In the convolution process over fitting problem may occur so there need to produce another methodology to address the issue.

- On the other side, there is no general mapping in the existing literature in the case of severity grading measurement. Some used binary DR classification while other used multi stage classification.

## 6. Conclusion & Future Work

In this literature, the existing researches are divided by the measurement of severity grading due to the requirement and efficient usage of the clinical practice and this study presented a state-of-the-art survey on the severity grading of DR detection. The primary purpose of DR detection is to early diagnosis and to prevent the disease from developing into its severe stage. To the date, Gulshan *et al.* [43] achieved highest performance of 0.99 AUC applying deep CNN method. However the scope still remains to understand what types of features is learned during the deep CNN training procedure. The advantage of deep CNN network is that the architecture can distinguish the lesions present in the image which are not visible even by the ophthalmologist and the performance measurement of the algorithm are compared to the provided accuracy by the ophthalmologist. Hence it is also necessary to investigate more how deep CNN achieve this. In a recent work, a comprehensive comparative review paper [61] has been presented for selecting appropriate Deep CNN architecture for practical implication. They concluded that there is no linear relationship between Deep CNN model complexity and accuracy and not all the models use their parameters with the same level of efficiency. They also stated that almost all models are capable of real time performance on high end GPU. In conclusion use of deep CNN architecture for severity grading shown highest performance but still this research is still new and in evolution. Limitation of deep learning based existing methods are that they require high end GPU implementation to get the desirable output, otherwise it will take long time to train the deep architecture [62]. Other scopes include:

a.  There is a need for more standardization of experimental evaluation of the existing methods in the case of severity grading. Some methods experimented based on binary classification while other used multi class for severity grading scale performance.

b.  It is also necessary to look into the downsampling procedure of the convolution level (of CNN based methods) which may lead to loss of high discriminating image feature.

c.  As another future work, hybrid architecture of deep CNN training network and more stronger Supervised Machine Learning (SVM) classification might lead to increase accuracy of the system as Huang & LeCun shown convolution network based architecture are good at learning invariant features but not always optimal for classification [63]. SVM is a learning technique from a training set of labeled examples created by expert solution.

d.  Lastly, existing approaches employed very deep CNN architectures for example GoogleNet by Lam *et al.* [45], ImageNet by Abramoff *et al.* [47] and Gulsan *et al.* [43]., which requires learning more than millions training parameters only during *fine-tuning* and *pre-trained* on a large scale datasets. This leads to a high-end resource bounded and computationally expensive technique. To overcome this problem as a future work, more lightweight All-ConvNet model can be proposed which will be more suitable and efficient.